\documentclass[prb,twocolumn,superscriptaddress,floatfix,showpacs]{revtex4}

\usepackage{amsmath}
\usepackage{amstext}
\usepackage{amssymb}
\usepackage{latexsym}
\usepackage[dvips]{graphicx}

\newcommand{\beq}{\begin{equation}}
\newcommand{\eeq}{\end{equation}}
\newcommand{\ket}[1]{\left\vert#1\right\rangle}
\newcommand{\Ham}{\mathcal H}
\newcommand{\barr}{\begin{eqnarray}}
\newcommand{\earr}{\end{eqnarray}}

\begin{document}

\title{Density of kinks after a sudden quench
  in the quantum Ising spin chain}

\author{Sei Suzuki}
\affiliation{Department of Physics and Mathematics, 
Aoyama Gakuin University, Fuchinobe, Sagamihara 229-8558, Japan}

\author{Davide Rossini}
\affiliation{International School for Advanced Studies (SISSA),
  Via Beirut 2-4, I-34014 Trieste, Italy}

\author{Giuseppe E. Santoro}
\affiliation{International School for Advanced Studies (SISSA),
  Via Beirut 2-4, I-34014 Trieste, Italy}
\affiliation{CNR-INFM Democritos National Simulation Center, 
  Via Beirut 2-4, I-34014 Trieste, Italy}
\affiliation{International Centre for Theoretical Physics (ICTP),
  I-34014 Trieste, Italy}

\date{\today}

\begin{abstract}

We investigate the time evolution of the density of kinks
in the spin-$1/2$ quantum Ising spin chain after a sudden quench in
the transverse field strength, and find that it relaxes to a value
which depends on the initial and the final values of the transverse
field, with an oscillating power-law decay.
We provide analytical estimates of the long-time behavior and
of the asymptotic value reached after complete relaxation,
and discuss the role of quantum criticality in the quench dynamics.
We show that, for a dynamics at the critical point,
the residual density of kinks after the quench
can be described by equilibrium statistical mechanics at
a finite temperature dictated by the energy of the state
after the quench. On the other hand, outside of criticality it
does not exhibit thermalization.

\end{abstract}

\pacs{75.40.Gb, 75.40.Cx, 75.10.Pq, 73.43.Nq}   

%75.40.Gb 	Dynamic properties (dynamic susceptibility, spin waves, spin diffusion, dynamic scaling, etc.) 
%75.40.Cx 	Static properties (order parameter, static susceptibility, heat capacities, critical exponents, etc.) 
%75.10.Pq 	Spin chain models 
%73.43.Nq 	Quantum phase transitions

\maketitle

%--------------------------------------------
\section{Introduction}
%--------------------------------------------

The experimental developments in the physics of ultracold
atomic gases that have been put forward in the last decade
opened up the possibility to
test some fundamental aspects of strongly interacting quantum
many-body systems~\cite{lewenstein07,bloch08}.
In particular, by loading cold atoms in optical lattices,
the implementation of Hubbard-like Hamiltonians~\cite{jaksch98}
or artificial spin chain models~\cite{duan03,jane03} has become possible.
The key features of these setups are a detailed microscopic knowledge
of the Hamiltonian, together with an extremely high accuracy in
controlling the system parameters with tunable external fields.
Moreover, due to the surprisingly long coherence times and to
small thermal fluctuations, the dynamics can be monitored
with very low dissipation and noise effects~\cite{bloch08}.
Recently emphasis has been put in the investigation of
out-of-equilibrium properties of strongly correlated systems.
We quote for example the spectacular observation of
a superfluid-to-Mott insulator quantum phase transition
in optical lattices~\cite{greiner02};
the formation of topological defects during a quench of trapped
atomic gases through a critical point~\cite{sadler06,weiler08};
the absence of thermalization in the non-equilibrium evolution
of an integrable Bose gas~\cite{kinoshita06}.

Such enormous experimental potentialities raised an increasing
theoretical interest in the study of the effects induced
by a sudden perturbation in such strongly correlated systems,
which goes beyond a mere debate of statistical physics
principles~\cite{igloi00,sengupta04,cazalilla06,calabrese06,rigol07,kollath07,manmana07,rigol08,cramer08,barthel08,gangardt08,kollar08,cramer08B,kollar08b,iucci09,calabrese09,rossini09,moeckel08,eckstein09,rigol09}.
Integrability is believed to play a crucial role in the relaxation
to the steady state, even if a comprehensive scenario is still lacking.
The nonequilibrium dynamics of a nonintegrable system
is expected to thermalize at the level of individual eigenstates~\cite{rigol08},
following the standard statistical mechanics
prescriptions~\cite{kollath07,manmana07,rigol08,eckstein09}.
On the contrary, for integrable models with nontrivial integrals
of motion the asymptotic equilibrium states
usually carry some memory of the initial conditions and are not the thermal
ones~\cite{sengupta04,cazalilla06,calabrese06,rigol07,cramer08,barthel08,gangardt08,kollar08b,kollar08,cramer08B,iucci09,calabrese09}.
In this case it is however possible to derive a statistical
prediction for the steady state in terms of a generalized Gibbs ensemble~\cite{rigol07}.
Several works tested the relevance of such distribution for different classes of
observables~\cite{cazalilla06,calabrese06,rigol07,kollar08,iucci09},
and found that it indeed works, provided the observables are sufficiently
uncorrelated with respect to the constants of motion~\cite{gangardt08,kollar08b}.

In the present study we consider a quantum quench of the one-dimensional
transverse field Ising chain; we investigate the time evolution of
the density of kinks that are generated after a quench in the field strength,
focusing on the thermalization properties of this observable and on the role
of the quantum phase transition in the quench dynamics.
Due to the complete integrability of the Ising model, observables are in general
not expected to follow the prescriptions of statistical mechanics.
As a matter of fact, the sensitivity to integrability can be traced
back to the property of locality of operators with respect to the
fermion quasiparticles that diagonalize the model in the continuum limit.
The Ising chain possesses two sectors of operators: one that is local
with respect to those particles, and another one that is non-local~\cite{zamolod91}.
Observables belonging to the non-local sector, as the two-point
correlation function of the order parameter, behave thermally~\cite{rossini09};
on the other hand, local observables in general do not.
Here indeed we find that the converged value of the density of kinks
after the quench is not equal to a thermal expectation value,
unless the system is quenched towards its critical point.

The paper is organized as follows. In Sec.~\ref{sec:model}
we describe our model and review some standard techniques
that are used to approach the quenched dynamics of the Ising chain.
In Sec.~\ref{sec:kinks} we define the density of kinks,
that is the quantity that we are going to analyze, and give
the prescription to evaluate it within the free fermion formalism.
We first focus on its behavior for the system at equilibrium,
both at zero and at finite temperatures (Sec.~\ref{sec:equil}),
and then concentrate on the quenched case at zero temperature
(Sec.~\ref{sec:quench}), where we analyze the asymptotic value
as well as the finite-time transient.
In Sec.~\ref{sec:nonthermal} we explicitly show that
the density of kinks does not exhibit a thermal behavior
outside criticality, while it thermalizes only at the critical point.
Finally in Sec.~\ref{sec:concl} we draw our conclusions.
In the appendix we derive explicit analytic expressions
for the density of kinks of the ground state (App.~\ref{app:kink_GS})
and after a quench (App.~\ref{app:kink_QU}).

%--------------------------------------------
\section{Model} \label{sec:model}
%--------------------------------------------

We consider the simplest nontrivial example of exactly
solvable one-dimensional quantum many-body systems, exhibiting a
quantum phase transition, that is the spin-$1/2$ quantum Ising
chain with ferromagnetic interactions~\cite{sachdev_book}.
This is characterized by the Hamiltonian
\beq
\Ham (\Gamma) =
- J \sum_j \left( \sigma^x_j \sigma^x_{j+1} + \Gamma \sigma^z_j \right) \,,
\label{eq:model}
\eeq
where $\sigma^\alpha_j$ ($\alpha = x,y,z$) are the Pauli matrices for
the $j$th spin, while the parameters $J$ and $\Gamma$ respectively
denote the nearest-neighbor antiferromagnetic exchange coupling
and the transverse field strength (hereafter we set $J=1$ as the
system's energy scale and take $\Gamma \geq 0$ without loss of generality;
we also use units of $\hbar = k_B = 1$).
In this paper we will assume periodic boundary conditions
for the Ising chain, and then take the limit $L \to \infty$,
in order to approach the thermodynamic limit.
We will therefore suppose that the sum in
Eq.~\eqref{eq:model} goes from $1$ to $L$,
with the rule $\sigma^\alpha_{L+1} \equiv \sigma^\alpha_1$
and with $L$ being the number of spins in the chain.
We assume an even number for $L$.
At zero temperature, the system is a quantum paramagnet
for $\Gamma>1$ and a ferromagnet with respect to the coupling direction
for $\Gamma<1$; these two phases are separated by a quantum critical point
at $\Gamma_c = 1$.

We choose to drive the system~\eqref{eq:model} out of equilibrium
by performing a sudden quench of the Hamiltonian parameter $\Gamma$:
we assume that the transverse magnetic field strength is suddenly
quenched from $\Gamma_0$ to $\Gamma$ at time $t=0$.
The system is also supposed to be initialized in 
the ground state $\ket{\Psi(\Gamma_0)}$ of $\Ham(\Gamma_0)$.
Due to the sudden variation of the field intensity,
at $t>0$ the state $\ket{\Psi(\Gamma_0)}$ will evolve according to
the new Hamiltonian $\Ham(\Gamma)$:
\beq
   \ket{\psi(t)} = U(t) \ket{\psi(0)}
   \equiv e^{- i \Ham(\Gamma) t} \ket{\Psi(\Gamma_0)} \, .
   \label{eq:psit}
\eeq
Due to the complete integrability of the Ising model,
it is generally possible to reduce the computational cost for evaluating
directly measurable quantities (such as dynamical correlation
functions or few-body observables) to a linear increase with the system
size $L$.
In some cases one can also derive an explicit analytic expression,
as we shall explain below.

%--------------------------------------------
\subsection{Statics}
%--------------------------------------------

The Ising Hamiltonian~\eqref{eq:model} can be exactly diagonalized by means
of a Jordan-Wigner transformation (JWT), followed by a Bogoliubov
rotation~\cite{lieb61,pfeuty70}.
First one represents the spins in terms of JW fermions,
which are defined by:
\beq
c_l \equiv \sigma_l^- \exp \bigg( i \pi
\sum_{j=1}^{l-1} \sigma^+_j \sigma^-_j \bigg) \, ,
\eeq
with $\sigma^\pm_j = (\sigma^x_j \pm i \sigma^y_j)/2$ raising and
lowering spin operators.
This leads to a Hamiltonian that is quadratic in the $c$-fermions.
Since the ground state has an even number of $c$-fermions,
and their parity is conserved during the evolution,
we will assume that the actual system state $\ket{\psi(t)}$ always
lies in the even $c$-fermionic Hilbert space sector.
This requires imposing antiperiodic boundary conditions for the fermions.

The Hamiltonian can then be easily handled by
switching to momentum representation:
%\beq
$c_k = \frac{1}{\sqrt{L}} \sum_j e^{-i j k} c_j$,
%\eeq
where the possible values of $k$ are fixed by the antiperiodic
boundary conditions, and are given by
$k = \pm \frac{\pi (2 n +1)}{L}$, with $n = 0,1, \ldots, \frac{L}{2}-1$.
Indeed, $\Ham$ decouples into a sum of independent terms,
each of them acting on the subspace ($-k, +k$).
This is finally diagonalized through a Bogoliubov transformation,
by introducing the new fermionic operators
\beq
\left\{ \begin{array}{lcl}
\gamma_k & = & (\bar{u}_k^\Gamma)^* \, c_k + (\bar{v}^\Gamma_k)^* \, c^\dagger_{-k}
\vspace*{0.5mm} \\
\gamma^\dagger_{-k} & = & -\bar{v}_k^\Gamma \, c_k + \bar{u}_k^\Gamma \, c^\dagger_{-k}
\end{array} \right. \eeq
where the coefficients
\beq
\bar{u}_k^\Gamma = \frac{\epsilon_k^\Gamma + a_k^\Gamma}
{\sqrt{2 \epsilon_k^\Gamma (\epsilon_k^\Gamma + a_k^\Gamma)}} \; , \quad
\bar{v}_k^\Gamma = \frac{i b_k}{\sqrt{2 \epsilon_k^\Gamma
(\epsilon_k^\Gamma + a_k^\Gamma)}} \, ,
\eeq
with $a_k^\Gamma = -2 (\cos k + \Gamma)$ and $b_k = 2 \sin k$,
while their dispersion is characterized by
\beq
\epsilon^\Gamma_k \equiv \sqrt{(a_k^\Gamma)^2 + (b_k^\Gamma)^2}
= 2 \sqrt{1 + \Gamma^2 + 2 \Gamma \cos k} \,  .
\label{eq:spectrum}
\eeq
The Ising model is then recasted into a free fermionic system,
where the Hamiltonian~\eqref{eq:model} can be written as:
\beq
\Ham(\Gamma) = \sum_{0 < k < \pi}
\epsilon_k^\Gamma \big( \gamma^\dagger_k \gamma_k
+ \gamma^\dagger_{-k} \gamma_{-k} -1 \big) \, .
\label{eq:hamdiag}
\eeq
The ground state $\ket{\Psi(\Gamma)}$ of $\Ham(\Gamma)$
is therefore the vacuum state for the
fermionic Bogoliubov quasiparticles $\gamma_k$:
\beq
\ket{\Psi(\Gamma)} = \prod_{0 < k < \pi} \gamma_k \gamma_{-k} \ket{0} \, ,
\label{eq:ground}
\eeq
where $\ket{0}$ is the vacuum state of $c$-fermions.

%--------------------------------------------
\subsection{Dynamics}
%--------------------------------------------

The dynamics of the Ising chain is conveniently described within the
Heisenberg representation~\cite{mccoy1}.
The Heisenberg representation of an operator ${\cal O}(t)$ is defined by
${\cal O}^H(t) = U^\dagger(t) \, {\cal O}(t) \, U(t)$, and it evolves
in time according to the equation of motion:
\beq
i \frac{d}{dt} {\cal O}^H(t) = U^\dagger(t) \left( \left[ {\cal O}(t), \Ham(t) \right] U(t)
+ i \frac{d}{dt} {\cal O}(t) \right) U(t) \,,
\eeq
$[\cdot, \cdot]$ denoting the commutator of two operators.

Since the Ising Hamiltonian is quadratic in the $c$-fermions,
the Heisenberg equations of motion for the operators $c_k^H(t)$
are linear and can be solved with a standard Bogoliubov-de Gennes approach.
Their solution is written according to
\beq
\left\{ \begin{array}{l}
c_k^{\phantom{\dagger} H}(t) = u_k(t) \, \gamma_k^0 - v^*_k(t) \, {\gamma^0_{-k}}^\dagger
\vspace*{1mm} \\
c_k^{\dagger H}(t) = v_k(t) \, \gamma_k^0 + u^*_k(t) \, {\gamma^0_{-k}}^\dagger
\end{array} \right.
\label{eq:HEISc}
\eeq
where $\gamma_k^0$ are the Bogoliubov operators that diagonalize
$\Ham(\Gamma_0)$ at the initial time, and
the coefficients $u_k(t), v_k(t)$ obey the time-dependent
Bogoliubov-de Gennes equations and are given by
\beq
\left( \begin{array}{c} u_k(t) \\ v_k(t) \end{array} \right) =
U_k^\Gamma \bigg( \begin{array}{cc} e^{-i \epsilon_k^\Gamma t} & 0 \\
0 & e^{i \epsilon_k^\Gamma t} \end{array} \bigg) U_k^{\Gamma \, \dagger}
\left( \begin{array}{c} u_k(0) \\ v_k(0) \end{array} \right)
\label{eq:BdG}
\eeq
with initial conditions
$u_k(0) = \bar{u}_k^{\Gamma_0}$, $v_k(0) = \bar{v}_k^{\Gamma_0}$ and
\beq
U_k^\Gamma = \left( \begin{array}{cc} \bar{u}_k^\Gamma & - (\bar{v}^\Gamma_k)^*
\vspace*{1mm} \\
\bar{v}_k^\Gamma & \phantom{-} (\bar{u}^\Gamma_k)^* \end{array} \right) \, .
\eeq

%--------------------------------------------
\section{Density of kinks} \label{sec:kinks}
%--------------------------------------------

The quantity we will analyze throughout this paper is the density of defects (or kinks),
defined by:
\beq
{\cal N} = \frac{1}{L} \sum_j \frac{1 - \sigma^x_j \sigma^x_{j+1}}{2} \,,
\eeq
where the sum goes from $j=1$ to $N$, for periodic boundaries.
Switching to the Heisenberg representation, one can compute the
expectation value of the density of kinks $\rho(t)$ at a certain time $t$.
In the thermodynamic limit, this is defined by:
\beq
\rho(t) = \lim_{L \to \infty} \langle \Psi(\Gamma_0) \vert
{\cal N}^H(t) \vert \Psi(\Gamma_0) \rangle .
\eeq
In order to evaluate this expectation value, one has first to express
${\cal N}^{\rm H}(t)$
in terms of $c$-fermions, and then write them as combinations of 
quasiparticles $\gamma_k^0$ that diagonalize the Hamiltonian before the quench,
using Eqs.~\eqref{eq:HEISc}.
In terms of the coefficients $u_k(t)$ and $v_k(t)$, the expectation value is written as
\barr
\rho(t) &=& \int_0^{\pi} \frac{dk}{2\pi}
\Big[ 1 - \big( |v_k(t)|^2 - |u_k(t)|^2 \big) \cos k \nonumber\\
 && \hspace*{0.5cm}
- i \big( u_k(t)v_k^{\ast}(t) - u_k^{\ast}(t)v_k(t) \big) \sin k
\Big] \;.
\earr
Substituting the solution of the Bogoliubov equations for $u_k(t)$ and
$v_k(t)$, Eq.~\eqref{eq:BdG}, one obtains the following integral
expression for the density of kinks at zero temperature:
\barr
\label{eq:rhoZeroInt}
\rho(t) & = & \int_0^\pi \frac{dk}{2 \pi} \bigg\{
1 - 2 \frac{1+\Gamma_0 \cos k}{\epsilon_k^{\Gamma_0}} \\ \nonumber
&& -8 \frac{\Gamma (\Gamma_0 - \Gamma) \sin^2 k}{(\epsilon_k^\Gamma)^2
\, \epsilon_k^{\Gamma_0}} \Big[ 1 - \cos \big( 2 \epsilon_k^\Gamma \, t
\big) \Big] \bigg\} \, .
\earr

The generalization to finite temperatures $T$ is quite straightforward;
however, concerning the case $T>0$, we are only interested
in the equilibrium situation, to which we restrict ourselves.
If the average is taken on the thermal state at
temperature $T$, 
the density of kinks at equilibrium is defined by
\beq
\rho_T^{\rm eq} = \lim_{L \to \infty} \frac{1}{Z} \, {\rm Tr}
\big[ e^{- \beta \Ham(\Gamma)} {\cal N} \big] \, ,
\eeq
where $e^{-\beta \Ham(\Gamma)}/Z$ is the canonical ensemble of the system
at a given temperature $T = 1/ \beta$, and
$Z = {\rm Tr} [e^{-\beta \Ham(\Gamma)}]$ is the partition function.
At finite temperatures,
one has to take into account the
possibility of having thermally 
excited quasiparticles, according to the Fermi distribution function
\beq
n_\mu(T) = \frac{1}{\exp(\beta \, \epsilon_\mu^\Gamma)+1} \, ,
\label{eq:fermi}
\eeq
where $\epsilon_\mu^\Gamma$ is the energy of the $\gamma_\mu$ quasiparticle.
The quasiparticle averages on the thermal state can thus be evaluated
in the same way as on the ground state, using
$\langle \gamma^\dagger_\mu \gamma^\dagger_\nu \rangle_T =
\langle \gamma_\mu \gamma_\nu \rangle_T = 0$ and
$\langle \gamma^\dagger_\mu \gamma_\nu \rangle_T =
\big( 1 - \langle \gamma_\mu \gamma^\dagger_\nu \rangle_T \big) \delta_{\mu, \nu}
= n_\mu(T) \, \delta_{\mu, \nu}$,
where $\langle \cdot \rangle_T$ denotes the thermal average and $\delta_{\mu, \nu}$
is the Kronecker delta.
Proceeding in a way analogous to the zero-temperature case,
one obtains this expression for $\rho_T^{\rm eq}$:
\barr
\rho_T^{\rm eq} &=& \int_0^{\pi} \frac{dk}{2\pi}
\bigg[ 1 - \big( 1-2 n_k(T) \big) \big( |\bar{v}_k^\Gamma|^2 - |\bar{u}_k^\Gamma|^2 \big) \cos k
\nonumber\\ && 
- i \big( 1-2 n_k(T) \big) \big( \bar{u}_k^\Gamma (\bar{v}_k^\Gamma)^{\ast}
- (\bar{u}_k^\Gamma)^{\ast} \bar{v}_k^\Gamma \big) \sin k \bigg] \;.
\label{eq:finiteT}
\earr

In the following, we will analyze in detail Eq.~\eqref{eq:rhoZeroInt}
and Eq.~\eqref{eq:finiteT}.

%--------------------------------------------
\section{Behavior at equilibrium} \label{sec:equil}
%--------------------------------------------

Let us now start focusing on the expectation value of the density of kinks
for the case in which system is time-independent, 
i.e. at equilibrium. In this situation the density of kinks is a
time-independent quantity, and is given by the expression in
Eq.~\eqref{eq:finiteT}.

%--------------------------------------------
\subsection{Zero temperature} \label{subsec:Eq_T0}
%--------------------------------------------

At zero temperature the system is frozen in the ground state $\ket{\Psi(\Gamma)}$,
that is the vacuum of $\gamma_k$ particles, Eq.~\eqref{eq:ground},
therefore the Fermi distribution function reduces to zero and the only
non zero expectation value is $\langle \gamma_\mu \gamma^\dagger_\mu \rangle = 1$.
In this case, the density of kinks can be straightforwardly obtained
by putting $\Gamma_0 = \Gamma$ in Eq.~\eqref{eq:rhoZeroInt}, and is given by:
\beq
\rho_0 = \rho_{T=0}^{\rm eq} = \int_0^\pi \frac{dk}{2 \pi} \left[ 1-
\frac{1 + \Gamma \cos k}{\sqrt{1+\Gamma^2 + 2 \Gamma \cos k}} \right] \, .
\label{eq:kink_T0}
\eeq

This expression can be computed analytically, and is written
in terms of complete elliptic integrals~\cite{abramowitz},
which are defined as follows:
\barr
 K(\lambda) &\equiv& \int_0^1 \frac{du}{\sqrt{(1-u^2)(1-\lambda u^2)}}
\, ,  \label{eq:elliptickK} \\ 
 \Pi(c,\lambda) &\equiv& \int_0^1 \frac{du}{(1 - c u^2)
  \sqrt{(1-u^2)(1 - \lambda u^2)}} \, . \label{eq:ellipticPi}
\earr
Following Appendix~\ref{app:kink_GS} one gets:
\beq
\label{eq:Kink_T0}
 \rho_0 = \frac{1}{2} - \left\{ \begin{array}{ll} \displaystyle
\frac{1 - \Gamma^2}{\pi}  \Pi\big( \Gamma^2,\Gamma^2 \big)
& {\rm for} \;\; \Gamma < 1
\vspace*{2mm} \\ \displaystyle
\frac{1}{\pi} & {\rm for} \;\; \Gamma = 1 \vspace*{2mm} \\ \displaystyle
\frac{\Gamma^2 - 1}{\pi\Gamma}
 \biggl\{ \Pi \Big( \frac{1}{\Gamma^2}, \frac{1}{\Gamma^2} \Big)
  - K \Big( \frac{1}{\Gamma^2} \Big) \biggr\}
& {\rm for} \;\; \Gamma > 1 \, .\\
\end{array} \right.
\eeq
The density of kinks at equilibrium and at zero temperature,
as a function of the field strength $\Gamma$, is plotted
in Fig.~\ref{fig:Kink_Tfin} with black solid curve.
As one can see, $\rho_0$ is a monotonic increasing function with $\Gamma$,
and its first derivative diverges at the critical point $\Gamma_c = 1$.

%--------------------------------------------
\subsection{Finite temperature}
%--------------------------------------------

The density of kinks at equilibrium at finite temperatures
is given by Eq.~\eqref{eq:finiteT}, and can be reexpressed 
in the following way:
\beq
\rho_T^{\rm eq} = \int_0^\pi \frac{dk}{2 \pi} \left[ 1- 
\frac{1 + \Gamma \cos k}{\sqrt{1+\Gamma^2 + 2 \Gamma \cos k}}
\big( 1 - 2 n_k(T) \big)  \right] \,.
\label{eq:Kink_T}
\eeq

\begin{figure}[!t]
  \begin{center}
    \includegraphics[scale=0.33]{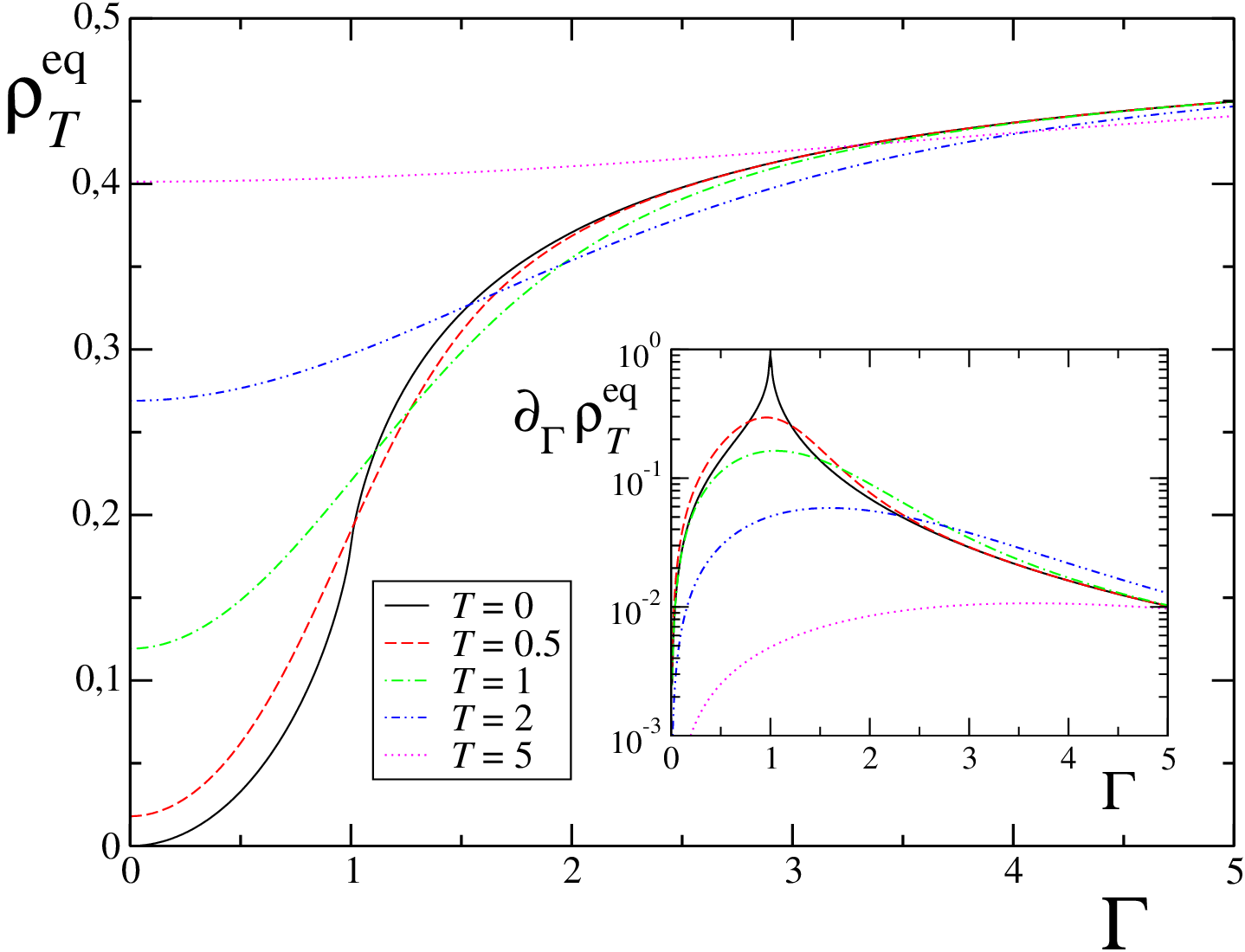}
    \caption{(color online). 
      Density of kinks $\rho_T^{\rm eq}$ for the thermal state
      at temperature $T$, as a function of the transverse field $\Gamma$.
      The black solid line denotes $\rho_0$, that is the equilibrium density of kinks
      in the ground state $\ket{\Psi(\Gamma)}$.
      Inset: first derivative of $\rho_T^{\rm eq}$ with respect to $\Gamma$.}
    \label{fig:Kink_Tfin}
  \end{center}
\end{figure}

Unlike the zero temperature case,
it is not possible to derive a simple analytic expression
for $\rho_T^{\rm eq}$, and one has to evaluate it numerically.
Results are shown in Fig.~\ref{fig:Kink_Tfin}, where we plot
$\rho_T^{\rm eq}$ as a function of the transverse field strength $\Gamma$.
As for the $T=0$ case, $\rho_T^{\rm eq}$ increases monotonically with $\Gamma$,
towards the maximum value $1/2$, that is reached
in the limit $\Gamma \to \infty$.
The singularity in its first derivative at $\Gamma_c$ disappears
for $T>0$ and is progressively smoothed out, as $T$ increases.

%%%%%%%%%%%%%%%%%%%%%%%%%%%%%
%%%%%%%%%%%%%%%%%%%%%%%%%%%%%

%--------------------------------------------
\section{Behavior after a quench} \label{sec:quench}
%--------------------------------------------

We now discuss in details the behavior of the density of
kinks at $T=0$ after a quantum quench in the transverse field.
As explained in Sec.~\ref{sec:model}, the system is supposed to be
initially in the ground state at $\Gamma_0$, and then evolved according
to the quenched Hamiltonian, following Eq.~\eqref{eq:psit}.
Therefore, concerning the non-equilibrium situation,
we specialize to the zero-temperature case,
and denote with $\rho_Q(t)$ the density of kinks.
This is explicitly given by Eq.~\eqref{eq:rhoZeroInt},
and can be written
as the sum of a time-independent part and a time-dependent term,
which decays in time and vanishes for asymptotically long times:
\beq
\rho_Q(t) \equiv \rho_Q^{(0)} + \rho_Q^{(1)} (t) 
\label{eq:kink_quench1}
\eeq
with $\rho_Q^{(1)} (t) \stackrel{t \to \infty}{\to} 0$.
In the rest of this section we will separately discuss
the behaviors of $\rho_Q^{(0)}$ and $\rho_Q^{(1)} (t)$.

%--------------------------------------------
\subsection{Asymptotic value}
%--------------------------------------------

Similar to the density of kinks at zero temperature, $\rho_0$,
the asymptotic value $\rho_Q^{(0)}$ can be expressed
in terms of complete elliptic integrals $ K(\lambda)$
and $\Pi(c,\lambda)$, Eqs.~\eqref{eq:elliptickK} and
\eqref{eq:ellipticPi}. 
We introduce the following notation for convenience:
\beq
 \varphi_1 \equiv \Gamma^2 - 2 \Gamma_0 \Gamma + 1 \; , \quad
 \varphi_2 \equiv \Gamma_0 \Gamma^2 + \Gamma_0 - 2 \Gamma \, .
\label{eq:phi1phi2}
\eeq
The expression of $\rho_Q^{(0)}$ is then given separately
for the following three cases with respect to the value of $\Gamma_0$,
as explained in Appendix~\ref{app:kink_QU}. \\
(i) Case $\Gamma_0 > 1$:
\barr
\label{eq:rho_Q_0_gt_1}
 \rho_Q^{(0)} &=& \frac{1}{2} - \frac{1}{2\pi}\left[
  \frac{2(\Gamma_0^2 - 1)(\Gamma_0 \Gamma - 1)}
  {\Gamma_0 \varphi_1} K \Big( \frac{1}{\Gamma_0^2} \Big) \right. \\
 && + \frac{( \Gamma_0 + \Gamma )(\Gamma_0^2 - 1)}{\Gamma_0^2} 
  \Pi \Big( \frac{1}{\Gamma_0^2}, \frac{1}{\Gamma_0^2} \Big) \nonumber\\
 && + \left. \frac{(\Gamma_0 - \Gamma)(\Gamma_0^2 - 1)(\Gamma^2 - 1)^2}
  {\Gamma_0 \varphi_1 \varphi_2 }
  \Pi \Big( \big( \frac{\varphi_1}{\varphi_2} \big)^2, \frac{1}{\Gamma_0^2} \Big)
\right] . \nonumber 
\earr
(ii) Case $\Gamma_0 < 1$:
\barr
\label{eq:rho_Q_0_lt_1}
 \rho_Q^{(0)} &=& \frac{1}{2} - \frac{1}{2\pi}\left[
   -\frac{2\Gamma(\Gamma_0-\Gamma)(1-\Gamma_0^2)}{\Gamma_0
   \varphi_2}K(\Gamma_0^2) \right. \\ 
 && + \frac{(\Gamma_0+\Gamma)(1-\Gamma_0^2)}{\Gamma_0}\Pi(\Gamma_0^2,
 \Gamma_0^2) \nonumber\\ 
 && + \left. \frac{(\Gamma_0-\Gamma)(1-\Gamma_0^2)(1-\Gamma^2)^2}
       {\varphi_1\varphi_2}
       \Pi \Big( \big( \frac{\varphi_2}{\varphi_1} \big)^2, \Gamma_0^2 \Big)
\right] . \nonumber
\earr
(iii) Case $\Gamma_0 = 1$:
\barr
\label{eq:rho_Q_0_eq_1}
 \rho_Q^{(0)} = \frac{1}{2} - \frac{1+\Gamma}{2\pi} - \frac{1 -
 \Gamma^2}{8\pi\sqrt{\Gamma}} \,
 \ln \left[ \frac{1 + \Gamma + 2\sqrt{\Gamma}}{1 + \Gamma - 2\sqrt{\Gamma}} 
\right] \;.
\earr

\begin{figure}[!t]
  \begin{center}
    \includegraphics[scale=0.31]{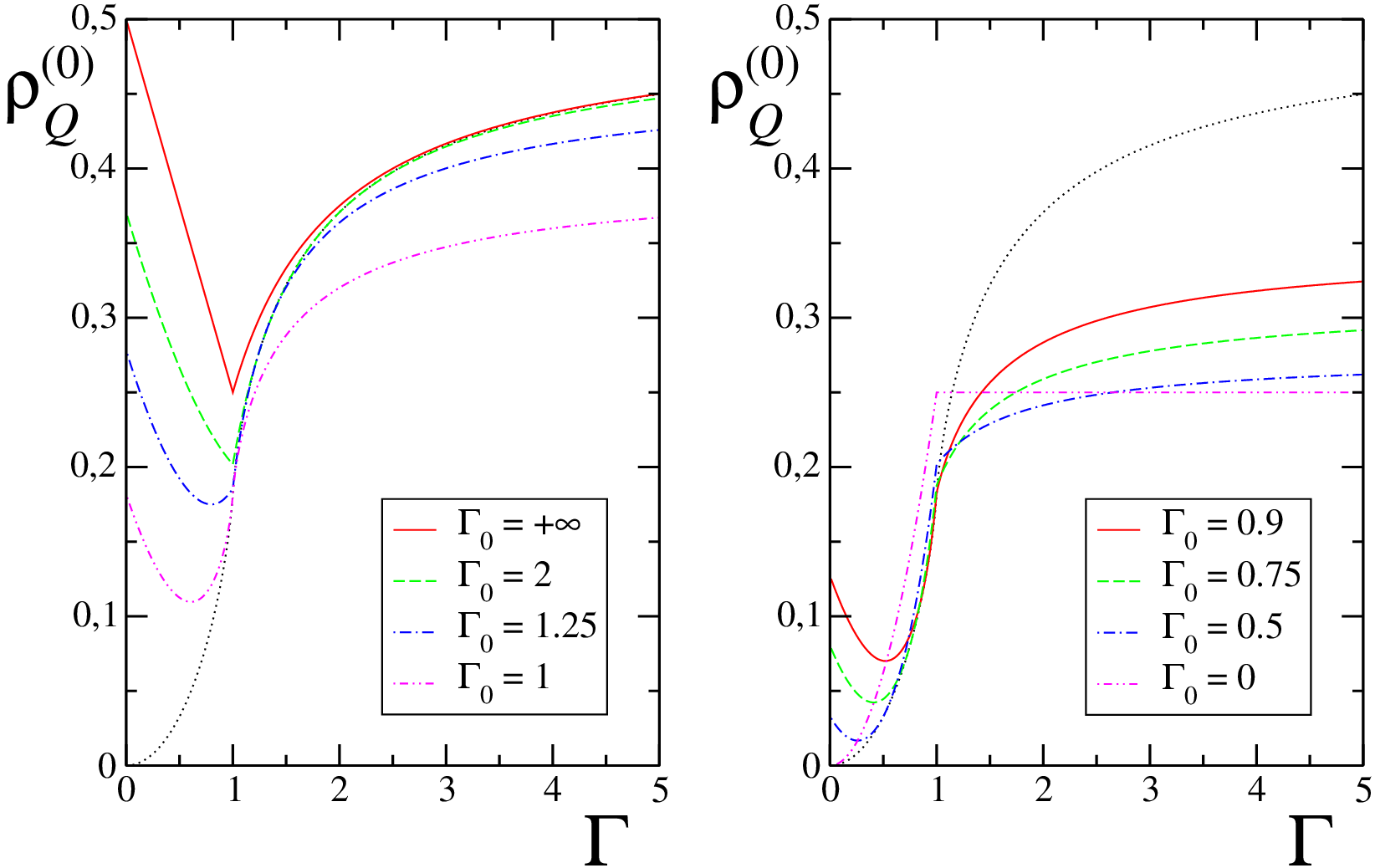}
    \caption{(color online). Asymptotic value $\rho_Q^{(0)}$
      of the density of kinks for $t \to \infty$ for a
      quench from $\Gamma_0 \geq 1$ (left panel) and
      from $\Gamma_0 < 1$ (right panel), as a function of
      the $\Gamma$ after the quench.
      The black dotted line shows the $T=0$ equilibrium density of kinks $\rho_0$.}
    \label{fig:Asynt_G}
  \end{center}
\end{figure}

In Fig.~\ref{fig:Asynt_G} we show the asymptotic value of the density
of kinks after the quench, as a function of the transverse field $\Gamma$.
We notice that, when the system after the quench is a paramagnet
($\Gamma > 1$), the density of kinks always decreases with
decreasing $\Gamma$, for any value of the initial field $\Gamma_0 >0$.
In that case, the shape of the curve is qualitatively the same
as that for the ground state value $\rho_0$, even if it has
a smaller gradient.
The situation is subtler for quenches ending in the ferromagnet,
where the behavior of $\rho_Q^{(0)}$ becomes non monotonic.
As a matter of fact, decreasing $\Gamma$, $\rho_Q^{(0)}$ has a
minimum at a certain value $\Gamma<1$ and then it increases
abruptly.
%Notice however that defects produced by the quench across
%the critical point from the paramagnet increase
%with the quench strength, and never attain the value $\rho_k^0$.

%%%%%%%%%%%%%%%%%%%%%%%%%%%%%
%%%%%%%%%%%%%%%%%%%%%%%%%%%%%

%--------------------------------------------
\subsection{Time dependent transient}
%--------------------------------------------

The relaxation of the density of kinks toward its asymptotic
value can be unveiled by an asymptotic expansion of the 
time-dependent part of Eq.~\eqref{eq:rhoZeroInt},
$\rho_Q^{(1)}(t)$, for large $t$.
As explained in Appendix~\ref{app:kink_QU},
an explicit expression for the asymptotic expansion can be
given separately for the following cases,
up to order $\mathcal{O}(t^{-2})$ in time.\\
(i) Case $\Gamma_0 \neq 1$ and $\Gamma\neq 1$:
\barr
 \label{eq:TimeDep1}
 \rho_Q^{(1)}(t) &=& \frac{\Gamma-\Gamma_0}{16\sqrt{2\pi
 \Gamma}\,t^{3/2}} \\
 && \times \biggl\{ \frac{1}{\sqrt{|1-\Gamma|}\,|1-\Gamma_0|}
 \cos \Big( 4|1-\Gamma|t -  \frac{\pi}{4} \Big)
   \nonumber \\
 && + \frac{1}{\sqrt{1+\Gamma}\,(1+\Gamma_0)} \cos \Big( 4(1+\Gamma)t +
 \frac{\pi}{4} \Big) \biggr\} \nonumber \\
 && + \mathcal{O}(t^{-2})
\, . \nonumber
\earr
(ii) Case $\Gamma_0 \neq 1$ and $\Gamma = 1$:
\beq
 \label{eq:TimeDep2}
 \rho_Q^{(1)}(t) = \frac{1-\Gamma_0}{32\sqrt{\pi}(1+\Gamma_0)\, t^{3/2}}
  \cos \Big( 8t + \frac{\pi}{4} \Big) + \mathcal{O}(t^{-2}) \, .
\eeq
(iii) Case $\Gamma_0 = 1$ and $\Gamma \neq 1$:
\barr
 \label{eq:TimeDep3}
 \rho_Q^{(1)}(t) &=& - \frac{1}{8\pi} \biggl\{
 \frac{1}{t}\sin \Big( 4(1-\Gamma)t \Big) \\
 && + \frac{\sqrt{\pi}(1-\Gamma)}
 {4\sqrt{2 \Gamma (1+\Gamma) }t^{3/2}}
\cos \Big( 4(1+\Gamma)t + \frac{\pi}{4} \Big)
 \biggr\} \nonumber \\
 && + \mathcal{O}(t^{-2})
\, . \nonumber
\earr

The decay of $\rho_Q^{(1)} (t)$ for large times is a power-law
with oscillations. This feature does not depend on whether the
quench is across the critical point or not.
If the system before the quench is not critical, $\Gamma_0\ne 1$,
the leading term decays as $t^{-3/2}$.
The oscillatory part of the leading term consists of two frequencies,
$4 (1 + \Gamma)$ and $4 \vert 1-\Gamma \vert$;
the second vanishes if the system is quenched to the critical point.
On the other hand, if the system before the quench is critical,
$\Gamma_0 = 1$, 
the leading term decays as  $t^{-1}$ with an oscillating term
of frequency $4 \vert 1-\Gamma \vert$.
These frequencies of the oscillation come from the modes with zero
group velocity, defined by $\partial \epsilon_k^{\Gamma}/\partial k = 0$.
Indeed, such modes are found from
Eq.~\eqref{eq:spectrum} to be $k = 0$ and $k = \pi$,
the latter of which disappears when $\Gamma=1$.
As it is seen from Eq.~\eqref{eq:rhoZeroInt}, 
the frequency corresponding to the mode $k$ is given by
$2\epsilon_k^{\Gamma}$. This leads to the frequencies 
$4(1+\Gamma)$ for $k=0$ and $4\vert 1-\Gamma \vert$ for $k=\pi$.

\begin{figure}[!t]
  \begin{center}
    \includegraphics[scale=0.34]{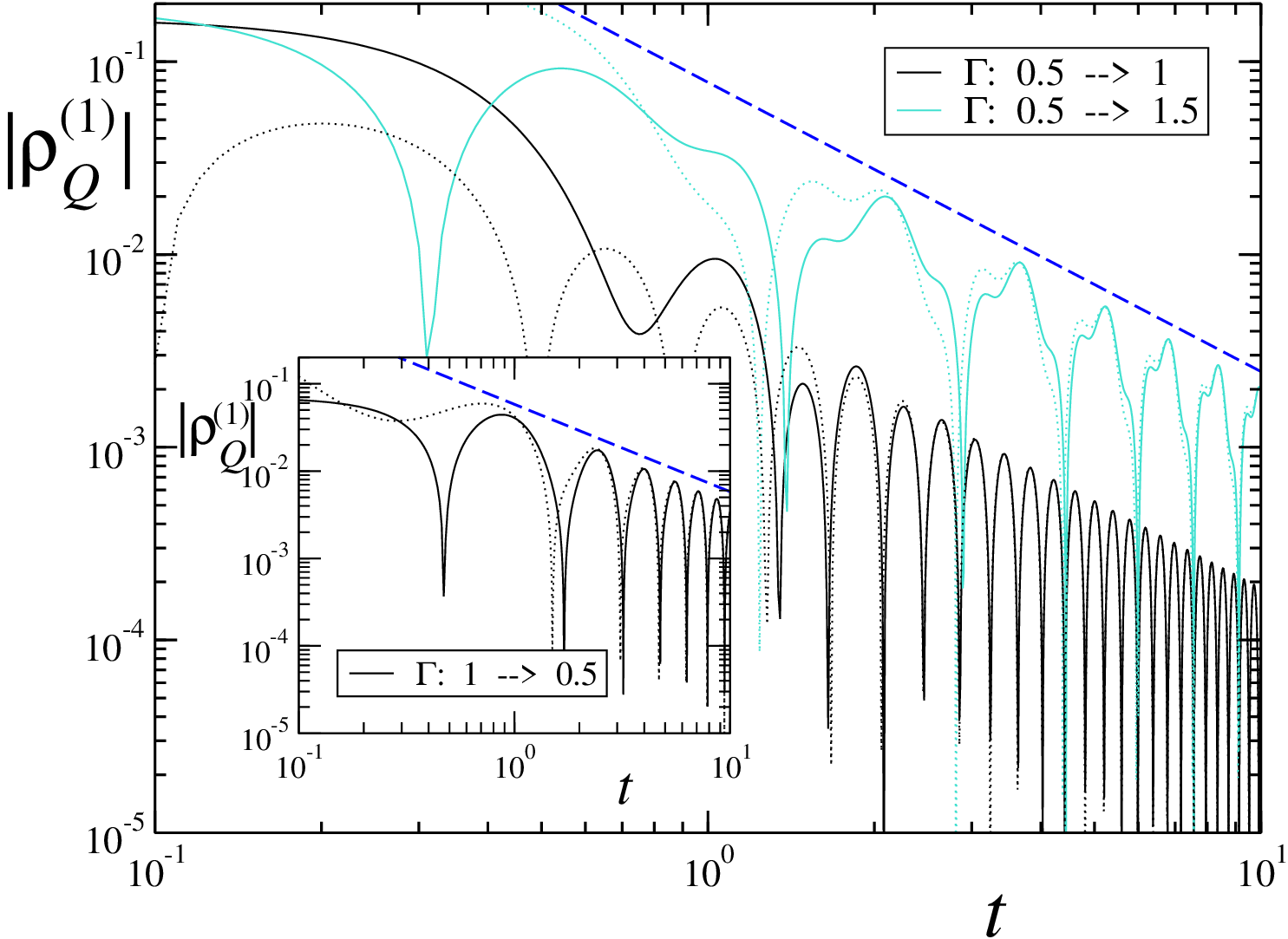}
    \caption{(color online). Absolute value of the time-dependent
      transient $\rho_Q^{(1)}$
      for the density of kinks after various types of quench.
      In the main panel we show two cases of quench from a
      non critical ground state, while in the inset we start
      from the critical point.
      Continuous curves are obtained from a numerical integration
      of Eq.~\eqref{eq:rhoZeroInt} while dotted curves are plots of
      the analytic expressions~\eqref{eq:TimeDep1}-\eqref{eq:TimeDep3}.
      The blue dashed lines denote the power-law behaviors
      $\rho_Q^{(1)} (t) \sim t^{-3/2}$ (main panel)
      and  $\rho_Q^{(1)} (t) \sim t^{-1}$ (inset), and
      are depicted as guidelines.}
    \label{fig:TimeDep_quench}
  \end{center}
\end{figure}

Figure~\ref{fig:TimeDep_quench} displays the behavior of $\rho_Q^{(1)} (t)$
for various types of quenches, obtained by numerical integration
of Eq.~\eqref{eq:rhoZeroInt} (solid curves), as well as
by the analytic asymptotic expressions in
Eqs.~\eqref{eq:TimeDep1}-\eqref{eq:TimeDep3} (dotted curves).
As it can be seen from the main panel, if the system before
the quench is not critical, the power-law exponent of the
decay is $t^{-3/2}$; notice also that, in the case where
the system is quenched outside criticality, the oscillations
are superpositions of different frequencies; if it is quenched
to the critical point, oscillations are more regular and
consist of just one frequency.
On the other hand, if the initial state is the ground state
critical, the decay is $t^{-1}$, as shown in the inset.

%%%%%%%%%%%%%%%%%%%%%%%%%%%%%
%%%%%%%%%%%%%%%%%%%%%%%%%%%%%

%--------------------------------------------
\section{Non-thermal behavior outside criticality} \label{sec:nonthermal}
%--------------------------------------------

We now concentrate on the asymptotic value $\rho_Q^{(0)}$ of the density
of kinks after a quench, and discuss the possibility to
track a behavior of this quantity in terms of an equilibrium
situation at a finite temperature~\cite{rossini09}.
In particular we would like to study if it is possible to define
an effective temperature for the quenched system (out of equilibrium)
in the most natural way so that, looking at $\rho_Q^{(0)}$,
the system behaves as if it was at equilibrium, at the same effective temperature.

Due to the quench, the ground state $\ket{\Psi(\Gamma_0)}$
of a given Hamiltonian $\Ham(\Gamma_0)$ becomes an
excited state for the quenched Hamiltonian $\Ham(\Gamma)$,
therefore it has a positive energy
$E_0 = \frac{1}{L} \langle \Psi(\Gamma_0) \vert \Ham(\Gamma) \vert \Psi(\Gamma_0) \rangle$,
as compared to the ground energy (we take all the energies normalized
per single site).
An effective temperature for the system out of equilibrium can
be defined by equating such energy $E_0$ to that of a fictitious
thermal state for the quenched Hamiltonian:
\beq
\langle \Ham(\Gamma) \rangle_{T_{\rm eff}} =
\int_0^\pi \frac{dk}{2 \pi} \, \epsilon_k^\Gamma
\Big[ n_k(T_{\rm eff}) + n_{-k}(T_{\rm eff}) - 1 \Big]
\label{eq:ETeff}
\eeq
where we used the representation of $\Ham(\Gamma)$ in terms
of free fermionic Bogoliubov quasiparticles, Eq.~\eqref{eq:hamdiag},
and $n_k(T)$ is given by Eq.~\eqref{eq:fermi}.
We therefore define the effective temperature by the implicit
equation\cite{rossini09}:
\beq
E_0 \equiv \langle \Ham(\Gamma) \rangle_{T_{\rm eff}} \, .
\label{eq:Teff}
\eeq
For a given pair of values $(\Gamma_0, \Gamma)$, this always
admits a solution;
we point out that, for a fixed $\Gamma$, there are two values of
$\Gamma_0$ corresponding to the same $T_{\rm eff}$,
one for $\Gamma_0 < \Gamma$ and the other for $\Gamma_0 > \Gamma$,
except at $T_{\rm eff} = 0$, which coincides with the static
case $\Gamma_0 = \Gamma$.

%%%%%%%%%%%%%%%%%%%%%%%%%%%%%
\begin{figure}[!t]
  \begin{center}
    \includegraphics[scale=0.33]{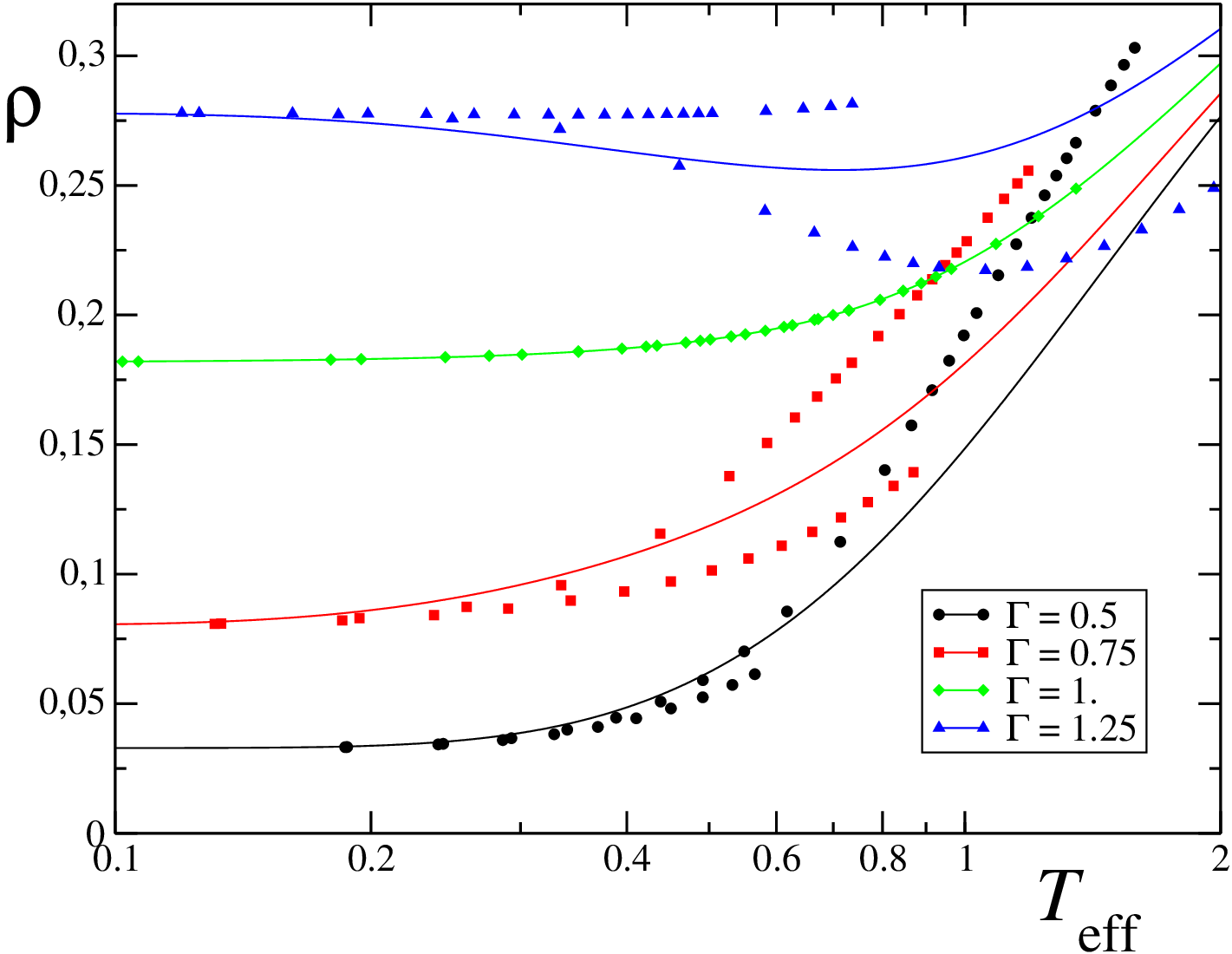}
    \caption{(color online). Asymptotic value of the density
      of kinks as a function of the effective temperature $T_{\rm eff}$.
      Different colors stand for various
      values of $\Gamma$, as explained in the caption.
      Symbols denote the density of kinks $\rho_Q^{(0)}$ after
      a quench; at each value of $\Gamma_0$ corresponds a different
      initial state, therefore a different effective temperature,
      as explained in the text.
      Straight lines denote the finite-temperature equilibrium values
      $\rho_{T = T_{\rm eff}}^{\rm eq}$.}
    \label{fig:Asyng_Temp}
  \end{center}
\end{figure}
%%%%%%%%%%%%%%%%%%%%%%%%%%%%%

We are now ready to perform a quantitative comparison
between the values of $\rho_Q^{(0)}$ after a quench and
$\rho_{T_{\rm eff}}^{\rm eq}$ at equilibrium, where $T_{\rm eff}$
for the out-of-equilibrium system is obtained from Eq.~\eqref{eq:Teff}.
This is done in Fig.~\ref{fig:Asyng_Temp}.
As one can see, outside criticality the two quantities are evidently
not related and behave in different ways.
On the other hand, at the critical point $\Gamma_c=1$
symbols (diamonds) perfectly follow the solid line (green data). 
We interpret this as a thermal behavior, in the sense that
the density of kinks after a quench to the critical point
is univocally determined by an effective temperature $T_{\rm eff}$
that depends only on the initial state energy after the quench,
according to Eq.~\eqref{eq:Teff}.
Remarkably such behavior is not found for a non-critical dynamics,
where fine details of the initial condition seem to be important.

The thermal behavior at criticality can be recovered analytically.
Indeed, in that case the density of kinks at equilibrium
with an effective temperature $T_{\rm eff}$ is given, by
substituting $\Gamma=\Gamma_c=1$ in Eq.~\eqref{eq:Kink_T}:
\barr
 \rho_{T_{\rm eff}}^{{\rm eq},\Gamma_c} &=& \int_0^{\pi} \frac{dk}{2 \pi}
 \left[ 1- \frac{1}{4}\epsilon_k^{\Gamma_c} \big[ 1 - 2 n_k(T_{\rm eff}) \big]  \right] \nonumber \\
 &=& \frac{1}{2} + \frac{ \langle \Ham(\Gamma) \rangle_{T_{\rm eff}} }{4} = \frac{1}{2} + \frac{E_0}{4} \;,
\label{eq:Teff_G1}
\earr
where $\epsilon_k^{\Gamma_c} = 2\sqrt{2 + 2\cos k}$ is the energy of the $\gamma_k$-fermion at criticality
($\Gamma=\Gamma_c=1$), the second equality follows from Eq.~\eqref{eq:ETeff} with $n_k(T) = n_{-k}(T)$, 
and the third from the definition of $T_{\rm eff}$ in Eq.~\eqref{eq:Teff}.
The energy $E_0$ of the state $|\Psi(\Gamma_0)\rangle$ is obtained directly from Eq.~\eqref{eq:hamdiag}, 
by evaluating $\langle \Psi(\Gamma_0) \vert \gamma^\dagger_k \gamma_k \vert \Psi(\Gamma_0) \rangle$, 
and is given by
\beq
 E_0 = - \int_0^{\pi} \frac{dk}{\pi}
  \frac{2(1 + \Gamma_0) (1 + \cos k)}{\epsilon_k^{\Gamma_0}} \;.
\eeq
On the other hand, the value of $\rho_Q^{(0)}$ for a quench at criticality reads, 
from Eqs.~\eqref{eq:rhoZeroInt} and \eqref{eq:kink_quench1}, 
\barr
\left. \rho_Q^{(0)}\right|_{\rm cr} = 
\int_0^{\pi} \frac{dk}{2\pi}  \bigg\{ 1 - 
 2\frac{1+\Gamma_0 \cos k}{\epsilon_k^{\Gamma_0}} 
  - 8\frac{(\Gamma_0 - 1) \sin^2 k} {(\epsilon_k^{\Gamma_c})^2 \, \epsilon_k^{\Gamma_0}} \bigg\}  \nonumber
\earr
which, after simple algebra, can be shown to reduce to $1/2+E_0/4$, hence obeying the rigorous equality 
\beq
\left. \rho_Q^{(0)}\right|_{\rm cr} = \rho_{T_{\rm eff}}^{{\rm eq},\Gamma_c} \;.
\eeq
%

%--------------------------------------------
\section{Conclusions} \label{sec:concl}
%--------------------------------------------

In this paper we have analyzed the behavior of the density of kinks
in the quantum Ising chain, both at equilibrium and after
a sudden quench in the transverse field.
At equilibrium the density of kinks is monotonically increasing with
the transverse field strength, and presents a divergence in its
first derivative at the critical point. This singularity is smoothed
out by finite temperature effects.
In the quenched situation, it exhibits a temporal decay down
to a residual value which depends on the initial
and the final values of the field.
The finite-time transient is characterized by an oscillating
power-law decay, whose rate depends on whether the initial
state of the system is critical or not.

We have also shown that, if the system is quenched towards
the critical point, the density of kinks in the out-of-equilibrium
situation can be described by equilibrium statistical mechanics, provided
an effective temperature is defined according to the energy
of the state after the quench.
We say that this observable exhibits a thermal behavior only
for a critical dynamics, while it does not thermalize for quenches to
non-critical phases.
These results should be regarded in the light of the discussion
in Ref.~\cite{rossini09}, where it was conjectured that only 
non-local operators in the fermion quasiparticles that diagonalize the
model in the continuum limit exhibit a general thermal behavior.
Indeed, we find non-thermal behavior for our density of kinks, 
which is a local operator in terms of fermions. 
However, the surprising thermal behavior at criticality should be better understood
in a more general framework.

\acknowledgments
We acknowledge fruitful discussions with R. Fazio, G. Mussardo and A. Silva.  
One of the authors (SS) is supported by Grant-in-Aid for
Scientific Research (Grant No. 20740225) from the Ministry of Education,
Culture, Sports, 
Science and Technology of Japan.

%+++++++++++++++++++++++++++++++++++++++++++++++
\appendix
%+++++++++++++++++++++++++++++++++++++++++++++++

%--------------------------------------------
\section{Density of kinks in the ground state} \label{app:kink_GS}
%--------------------------------------------

In this appendix we derive an analytic expression for the expectation
value of the $T=0$ density of kinks $\rho_0$ with respect to the system
ground state $\ket{\Psi(\Gamma)}$, in the thermodynamic limit.

As explained in Sec.~\ref{subsec:Eq_T0}, the density of kinks for
the ground state is given by Eq.~\eqref{eq:kink_T0}.
Applying the variable transformation $x = \cos k$ we get
\beq
 \rho_0 = \frac{1}{2} - \frac{1}{2\pi} \int_{-1}^{1}
  \frac{dx}{\sqrt{1-x^2}}\frac{1 + \Gamma x}{\sqrt{1 + \Gamma^2 + 2
  \Gamma x}} .
\label{eq:rho-k0-int}
\eeq
In order to evaluate the integral, we consider 
the following three cases with respect to $\Gamma$: 
\\
i) Case $\Gamma > 1$. We perform the change of variable
\[
 x = - \frac{\Gamma u + 1}{u + \Gamma} \, ,
\]
so that the integral in r.h.s. of Eq.~\eqref{eq:rho-k0-int} becomes
\barr
 &&\int_{-1}^1 \frac{dx}{\sqrt{1 - x^2}}
  \frac{1 + \Gamma x}{\sqrt{1 + \Gamma^2 + 2 \Gamma x}}  \nonumber \\
 && = \int_{-1}^1 \frac{du}{\Gamma} 
  \frac{\Gamma^2 - 1}{\sqrt{(1 - u^2)(1 - u^2/\Gamma^2)}}
  \biggl( -1 + \frac{1}{1 - u^2 / \Gamma^2} \biggr) \nonumber \\
 && = 2\frac{\Gamma^2 - 1}{\Gamma}\biggl\{ - K \Big( \frac{1}{\Gamma^2} \Big)
  + \Pi \Big( \frac{1}{\Gamma^2}, \frac{1}{\Gamma^2} \Big) \biggr\}, \nonumber
\earr
where complete elliptic integrals are defined by
Eqs.~\eqref{eq:elliptickK}-\eqref{eq:ellipticPi}.
Therefore we obtain:
\beq
 \rho_0 = \frac{1}{2} - \frac{\Gamma^2 - 1}{\pi\Gamma}
 \biggl\{ \Pi \Big( \frac{1}{\Gamma^2}, \frac{1}{\Gamma^2} \Big)
  - K \Big( \frac{1}{\Gamma^2} \Big) \biggr\} .
\eeq
ii) Case $\Gamma < 1$. From the change of variable
\[
 x = - \frac{u + \Gamma}{\Gamma u + 1} ,
\]
it follows that:
\barr
 && \int_{-1}^1 \frac{dx}{\sqrt{1 - x^2}}
  \frac{1 + \Gamma x}{\sqrt{1 + \Gamma^2 + 2 \Gamma x}}  \nonumber\\
 && = \int_{-1}^1 du \frac{1}{\sqrt{(1 - u^2)(1 - \Gamma^2 u^2)}}
  \frac{1 - \Gamma^2}{1 - \Gamma^2 u^2} \nonumber \\
 &&= 2 (1 - \Gamma^2) \, \Pi(\Gamma^2, \Gamma^2) . \nonumber
\earr
Therefore:
\beq
 \rho_0 = \frac{1}{2} - \frac{1 - \Gamma^2}{\pi} \,
 \Pi(\Gamma^2,\Gamma^2) .
\eeq
iii) Case $\Gamma = 1$.
The integral in Eq.~\eqref{eq:rho-k0-int} reduces to 
\[
 \frac{1}{\sqrt{2}}\int_{-1}^1 \frac{dx}{\sqrt{1 - x}} = 2 .
\]
Therefore: 
\beq
 \rho_0 = \frac{1}{2} - \frac{1}{\pi} .
\eeq

%--------------------------------------------
\section{Analytic calculation of the density of kinks after a quench}
\label{app:kink_QU}
%--------------------------------------------

We derive here analytic expressions for the density of kinks
after a quench $\rho_Q(t)$. We consider both the time-independent
part $\rho_Q^{(0)}$ and the time-dependent transient $\rho_Q^{(1)}(t)$.

%--------------------------------------------
\subsection{Asymptotic value}
%--------------------------------------------

The time-independent part $\rho_Q^{(0)}$ of the density of kinks
after a quench is written in integral form, starting from 
Eq.~\eqref{eq:rhoZeroInt}, as
\barr
 \rho_Q^{(0)} &=& \int_0^{\pi} \frac{dk}{2\pi}  \bigg\{ 1 - 
 2\frac{1+\Gamma_0 \cos k}{\epsilon_k^{\Gamma_0}} 
  -  8\frac{\Gamma (\Gamma_0 - \Gamma) \sin^2 k}
 {(\epsilon_k^{\Gamma})^2 \, \epsilon_k^{\Gamma_0}} \bigg\}  \nonumber\\
 &=& \frac{1}{2} + I \;. \label{eq:rho-k0-I}
\earr
Applying a variable transformation by $x = \cos k$ and
substituting 
$\epsilon_k^{\Gamma} = 2\sqrt{1 + \Gamma^2 + 2 \Gamma \cos k}$, 
the integral $I$ is explicitly written as:
\barr
 I &=& - \frac{1}{2\pi} \int_{-1}^1 \frac{dx}{\sqrt{1 - x^2}}
 \left\{ \frac{1 + \Gamma_0 x}{\sqrt{1 + \Gamma_0^2 + 2 \Gamma_0 x}}
 \right. \nonumber \\
 & & \left. +  \frac{\Gamma (\Gamma_0 - \Gamma)(1 - x^2)}
 {(1 + \Gamma^2 + 2 \Gamma x)\sqrt{1 + \Gamma_0^2 + 2 \Gamma_0 x}}
\right\} . \label{eq:AsymIntegral}
\earr
In order to evaluate the integral, we separately consider
three cases with respect to $\Gamma_0$:\\
i) Case $\Gamma_0 > 1$. We apply a variable transformation
\[
 x = - \frac{\Gamma_0 u + 1}{u + \Gamma_0} .
\]
Then Eq.\eqref{eq:AsymIntegral} is arranged into
\barr
 I &=& - \frac{1}{2\pi}\int_{-1}^1 \frac{du}{\Gamma_0^2}
  \frac{1}{\sqrt{(1-u^2)(1-u^2/\Gamma_0^2)}} \nonumber\\
 && \times \left\{
    \frac{\Gamma_0(\Gamma_0^2-1)(\Gamma_0\Gamma - 1)}
    {\varphi_1}  + \frac{(\Gamma_0+\Gamma)(\Gamma_0^2-1)}{2(1 -
    u^2/\Gamma_0^2)}  \right. \nonumber\\
 && + \left. \frac{\Gamma_0(\Gamma_0-\Gamma)(\Gamma_0^2-1)(\Gamma^2-1)^2}
  {2\varphi_1\varphi_2 ( 1 - \varphi_1^2u^2/
  \varphi_2^2 )} \right\} , \nonumber
\earr
where we remark the notation defined by Eq.~\eqref{eq:phi1phi2}.
Using the definition of the elliptic integrals, Eqs.~\eqref{eq:elliptickK}
and \eqref{eq:ellipticPi}, we reduce it to
\barr
 I &=&  - \frac{1}{2\pi}
  \left\{\frac{2 (\Gamma_0^2-1)(\Gamma_0\Gamma-1)}
  {\Gamma_0 \varphi_1}K \Big( \frac{1}{\Gamma_0^2} \Big) \right. \\
 &&  + \left. 
   \frac{(\Gamma_0+\Gamma)(\Gamma_0^2-1)}{\Gamma_0^2}\Pi \Big( \frac{1}{\Gamma_0^2}, 
   \frac{1}{\Gamma_0^2} \Big) \right.\nonumber\\
 && + \left. \frac{(\Gamma_0-\Gamma)(\Gamma_0^2-1)(\Gamma^2-1)^2}
   {\Gamma_0 \varphi_1\varphi_2}
   \Pi \Big( \big( \frac{\varphi_1}{\varphi_2} \big)^2 ,
   \frac{1}{\Gamma_0^2} \Big) \right\} . \nonumber
\earr
Hence, from Eq.~\eqref{eq:rho-k0-I}, we get the analytic expression
for $\rho_Q^{(0)}$ in Eq.~\ref{eq:rho_Q_0_gt_1}.\\
ii) Case $0 \leq \Gamma_0 < 1$. We perform a variable transformation
by
\[
 x = - \frac{u + \Gamma_0}{\Gamma_0 u + 1} .
\]
Using the new variable $u$, Eq.~\eqref{eq:AsymIntegral} is written
as
\barr
 I &=& - \frac{1}{2\pi} \int_{-1}^1 
  \frac{du}{\sqrt{(1-u^2)(1-\Gamma_0^2u^2)}} \nonumber \\
 && \times \left\{ - \frac{\Gamma(\Gamma_0-\Gamma)(1-\Gamma_0^2)}
  {\Gamma_0 \varphi_2} 
  + \frac{\Gamma_0 +
  \Gamma}{2\Gamma_0}\frac{1-\Gamma_0^2}{1-\Gamma_0^2u^2}
  \right. \nonumber \\ 
 && + \left. \frac{(\Gamma_0-\Gamma)(1-\Gamma_0^2)(1-\Gamma^2)^2}
  {2 \varphi_1\varphi_2 ( 1 - \varphi_2^2 u^2 / \varphi_1^2 )} \right\} .
       \nonumber
\earr
The integrals above are expressed in terms of complete elliptic
integrals to yield
\barr
 I &=& - \frac{1}{2\pi}\left\{
   -\frac{2\Gamma(\Gamma_0-\Gamma)(1-\Gamma_0^2)}{\Gamma_0
   \varphi_2}K(\Gamma_0^2) \right. \\
 && + \frac{(\Gamma_0+\Gamma)(1-\Gamma_0^2)}{\Gamma_0}\Pi( \Gamma_0^2,
   \Gamma_0^2 ) \nonumber\\ 
 && + \left. \frac{(\Gamma_0-\Gamma)(1-\Gamma_0^2)(1-\Gamma^2)^2}
       {\varphi_1\varphi_2}
       \Pi \Big( \big( \frac{\varphi_2}{\varphi_1} \big)^2, \Gamma_0^2 \Big) \right\}
       . \nonumber
\earr
Therefore, taking Eq.~\eqref{eq:rho-k0-I} into account,
one obtains the analytic expression for $\rho_Q^{(0)}$ in Eq.~\ref{eq:rho_Q_0_lt_1}.\\
iii) Case $\Gamma_0 = 1$. Eq.~\eqref{eq:AsymIntegral} is simplified into 
\[
 I = - \frac{1}{2\pi}\int_{-1}^1 \frac{dx}{\sqrt{1-x}}
 \left\{\frac{1}{\sqrt{2}} + \frac{\Gamma(1 - \Gamma)}{\sqrt{2}}
 \frac{1-x}{1 + \Gamma^2 + 2 \Gamma x}\right\} \;.
\]
The first term in r.h.s. yields $-1/\pi$. Regarding the second
term, we apply a variable transformation: $t = \sqrt{2\Gamma(1-x)}$.
Then integral is carried out and yields
\[
 - \frac{\Gamma(1-\Gamma)}{2\pi\sqrt{2}(2\Gamma)^{3/2}}
 \left\{ - 4\sqrt{\Gamma} + (1 + \Gamma) \, \ln
 \left[ \frac{1 + \Gamma+ 2\sqrt{\Gamma}}{1 + \Gamma - 2 \sqrt{\Gamma}}\right]
   \right\}  \;. 
\]
Summing the two terms, $I$ is written as
\[
 I = - \frac{1 + \Gamma}{2 \pi} - \frac{1 - \Gamma^2}{8\pi\sqrt{\Gamma}} \,
 \ln \left[ \frac{1 + \Gamma + 2\sqrt{\Gamma}}{1 + \Gamma - 2\sqrt{\Gamma}} \right] \;.
\]
From Eq.~\eqref{eq:rho-k0-I}, one obtains the analytic
expression for $\rho_Q^{(0)}$ in Eq.~\ref{eq:rho_Q_0_eq_1}.

%--------------------------------------------
\subsection{Time-dependent transient}
%--------------------------------------------

Following Eq.~\eqref{eq:rhoZeroInt}, the time-dependent part
$\rho_Q^{(1)}(t)$ of the density of kinks after a quench is written as
\[
 \rho_Q^{(1)}(t) = 8 \int_0^{\pi}\frac{dk}{2\pi}
 \frac{\Gamma(\Gamma_0 - \Gamma)\sin^2 k}{(\epsilon_k^{\Gamma})^2
 \epsilon_k^{\Gamma_0}}\cos(2\epsilon_k^{\Gamma} t) \;.
\]
Applying a variable transformation by 
$u = \frac{\epsilon_k^{\Gamma}}{2(\Gamma + 1)}$, 
this equation is arranged into
\barr
  \label{eq:rho_Q1}
 \rho_Q^{(1)}(t) &=& \frac{(\Gamma_0-\Gamma)(1+\Gamma)}
 {4\pi\sqrt{\Gamma_0\Gamma}}  \\
 && \times\int_{u_0}^1
  \frac{du}{u}\frac{\sqrt{(u^2 - u_0^2)(1-u^2)}}{\sqrt{u^2 - \psi}}
  \cos (t' u) , \nonumber
\earr
where we defined $u_0 = |1 - \Gamma|/(1 + \Gamma)$, $t' = 4(1 + \Gamma)t$,
and 
$\psi = (\Gamma_0 - \Gamma) (1 - \Gamma_0\Gamma)/\{\Gamma_0
(1+\Gamma)^2\}$. 
Since $\psi$ as a function of $\Gamma_0$ has a derivative
$\partial \psi/ \partial \Gamma_0 = \Gamma(1 - \Gamma_0^2)/\Gamma_0^2$,
it is monotonically increasing with $\Gamma_0$ when $0 < \Gamma_0 < 1$,
while it decreases when $\Gamma_0 > 1$. The maximum of $\psi$ 
for a given $\Gamma$ is found at $\Gamma_0 = 1$ and its value is
$(1-\Gamma)^2/(1+\Gamma)^2=u_0^2$.
The variable $\psi$ can be either positive or negative, depending on 
$\Gamma$ and $\Gamma_0$. \\
%Indeed, $\psi > 0$ when 
%$\frac{1}{\Gamma} < \Gamma_0 < \Gamma$ with $\Gamma > 1$ 
%and when $\Gamma < \Gamma_0 < \frac{1}{\Gamma}$
%with $\Gamma < 1$, 
%while $\psi < 0$ when $\Gamma_0 < \frac{1}{\Gamma}$
%or $\Gamma_0 > \Gamma$ with $\Gamma > 1$ and when
%$\Gamma_0 < \Gamma$ or $\Gamma_0 > \frac{1}{\Gamma}$ with
%$\Gamma < 1$.
%
i) Case $\Gamma_0 \neq 1$ and $\Gamma \neq 1$.
We introduce a number $a$ that satisfies
\[
 0 < a < \min\{u_0 - \textrm{Re}(\sqrt{\psi}), 1 - u_0\} ,
\]
where we remark that $u_0 - \textrm{Re}(\sqrt{\psi})$ is larger than zero,
since $0 < u_0 < 1$ and $\psi < u_0^2$. We define
\barr
 I_1(t') \hspace*{-1mm} &=& \hspace*{-1mm} \int_{u_0}^{u_0+a}  
 \frac{du}{u}\frac{\sqrt{(u^2 - u_0^2)(1-u^2)}}{\sqrt{u^2 - \psi}}
  \cos (t' u) ,
  \label{eq:DefI1} \\
 I_2(t') \hspace*{-1mm} &=& \hspace*{-1mm} \int_{u_0+a}^{1}  
 \frac{du}{u}\frac{\sqrt{(u^2 - u_0^2)(1-u^2)}}{\sqrt{u^2 - \psi}}
  \cos (t' u) .
  \label{eq:DefI2}
\earr
We first consider the asymptotic behavior of $I_1(t')$ for large $t'$.
Shifting the variable as $u' = u - u_0$,
the integral $I_1(t')$ is arranged into
\barr
 I_1(t') &=& \left[\frac{2u_0(1-u_0^2)}{u_0^2(u_0^2-\psi)}\right]^{1/2} 
 \int_0^a du' \cos \big[ t'(u'+u_0) \big] \nonumber \\
 && \times 
  \left[\frac{(1+\frac{u'}{2u_0})u'(1-\frac{u'}{1-u_0})
   (1+\frac{u'}{1+u_0})}{(1 + \frac{u'}{u_0})^2
   (1+\frac{u'}{u_0+\sqrt{\psi}})
   (1+\frac{u'}{u_0-\sqrt{\psi}})}
   \right]^{1/2} . 
   \label{eq:I1a}
\earr
Since for $0\leq u' \leq a$
\[
 \frac{u'}{2u_0}, ~\frac{u'}{1-u_0}, ~
 \frac{u'}{1+u_0}, ~ \frac{u'}{u_0}, ~
 \Bigl|\frac{u'}{u_0+\sqrt{\psi}}\Bigr|, ~ \mbox{and} ~
 \Bigl|\frac{u'}{u_0-\sqrt{\psi}}\Bigr|
\]
are non-negative and less than unity,
one can expand the integrand term inside the square brackets
into a power series of the type
$\left[ \; \cdot \; \right]^{1/2} = \sqrt{u'}(1 + a_1u' + a_2u'^2 + \cdots)$.
Applying this power series expansion and using asymptotic expansions
of the Fresnel's type integrals~\cite{abramowitz} to Eq.~\eqref{eq:I1a},
one obtains:
\barr
 I_1(t) &=& \left[\frac{2u_0(1-u_0^2)}{u_0^2(u_0^2-\psi)}\right]^{1/2} 
  \bigg\{ \frac{\sin [t' (u_0+a)]}{t'}  \nonumber \\
 && \times \sqrt{a} (1 + a_1 a + a_2 a^2 + \cdots) \nonumber \\
 && - \frac{\sqrt{\pi}}{2t'^{3/2}}
 \cos\Bigl[t'u_0 - \frac{\pi}{4}\Bigr] 
  +  \mathcal{O}(t'^{-2}) \bigg\}
  \nonumber
\earr
As a final result, after resumming the power series of $a$, the
integral $I_1(t)$ is evaluated as
\barr
 I_1(t) &=& \left[\frac{(2u_0+a)a\{1-(u_0+a)^2\}}
 {(u_0+a)^2\{(u_0+a)^2-\psi\}}\right]^{1/2}\frac{\sin [t'(u_0+a)]}{t'} 
 \nonumber\\
 && - \left[\frac{2u_0(1-u_0^2)}{u_0^2(u_0^2-\psi)}\right]^{1/2}
  \frac{\sqrt{\pi}}{2 t'^{3/2}} \cos \Bigl[t'u_0 - \frac{\pi}{4}\Bigr]
  \nonumber \\
 && + \mathcal{O}(t'^{-2}) .
  \label{eq:I1b}
\earr

We next consider $I_2(t)$. Transforming the variable by $u' = 1 - u$,
Eq.~\eqref{eq:DefI2} is arranged into
\barr
 I_2(t) &=& \left[\frac{2(1-u_0^2)}{1-\psi}\right]^{1/2} 
 \int_0^{1-u_0-a} du' \cos \big[ t'(1-u') \big] \nonumber \\
 && \times
  \left[\frac{(1-\frac{u'}{1+u_0})(1-\frac{u'}{1-u_0})u'(1-\frac{u'}{2})}
   {(1-u')^2(1-\frac{u'}{1+\sqrt{\psi}})(1-\frac{u'}{1-\sqrt{\psi}})}\right]^{1/2}
   . \nonumber
\earr
Similarly to the evaluation of $I_1(t)$, one can define 
power series of $u'$ for the term inside the square brackets:
$\left[ \; \cdot \; \right]^{1/2} = \sqrt{u'}(1 + b_1u' + b_2u'^2 + \cdots)$.
Using asymptotic expansions of the Fresnel's integrals~\cite{abramowitz},
the integral is evaluated asymptotically as
\barr
 I_2(t) &=& \left[\frac{2(1-u_0^2)}{1-\psi}\right]^{1/2} \bigg\{
 - \frac{\sin \big[ t' (u_0 + a) \big]}{t'} \sqrt{1-u_0-a}   \nonumber\\
 && \times 
   (1 + b_1(1-u_0-a) + b_2(1-u_0-a)^2 + \cdots) 
 \nonumber\\
 &&- \frac{\sqrt{\pi}}{2t'^{3/2}} \cos \Bigl[t' + \frac{\pi}{4}\Bigr]+ 
  \mathcal{O}(t'^{-2}) \bigg\} .
  \nonumber
\earr
Finally, returning the power series into an original function, 
one obtains
\barr
 I_2(t) \hspace*{-1mm} & = \hspace*{-1mm} &
- \left[\frac{ (2u_0+a) a \{ 1-(u_0+a)^2 \}}{ (u_0+a)^2 
\{ (u_0+a)^2 - \psi \} }\right]^{1/2} \! \frac{\sin \big[ t'(u_0+a) \big]}{t'} 
 \nonumber\\
 && - \left[\frac{2(1-u_0^2)}{1-\psi}\right]^{1/2} \!
  \frac{\sqrt{\pi}}{2 t'^{3/2}} \cos \Bigl[t'+\frac{\pi}{4}\Bigr] 
  \nonumber\\
 && + \mathcal{O}(t'^{-2}) .
\label{eq:I2a}
\earr
Looking at Eqs.~\eqref{eq:I1b} and~\eqref{eq:I2a}, one finds that
the first terms in r.h.s. of both equations cancel in $I_1(t)+I_2(t)$.
After expressing $u_0$, $\psi$ and $t'$ in terms of $\Gamma$,
$\Gamma_0$ and $t$, from Eq.~\eqref{eq:rho_Q1} we arrive
at the following expression:
\barr
 \rho_{Q}^{(1)}(t) &=& \frac{\Gamma-\Gamma_0}{16\sqrt{2\pi
 \Gamma}\,t^{3/2}} \label{eq:rho_Q1-1-final} \\
 && \times \biggl\{ \frac{1}{\sqrt{|1-\Gamma|}\,|1-\Gamma_0|}
 \cos \left[ 4|1-\Gamma|t - \frac{\pi}{4}\right]
   \nonumber\\
 && +  \frac{1}{\sqrt{1+\Gamma}\,(1+\Gamma_0)}
 \cos \left[4(1+\Gamma)t + \frac{\pi}{4}\right] \biggr\} \nonumber\\
 && + \mathcal{O}(t^{-2}) . \nonumber
\earr
ii) Case $\Gamma_0 \neq 1$ and $\Gamma = 1$.
Notice that $u_0 = 0$ and $\psi = - (1-\Gamma_0)^2/4\Gamma_0 < 0$.
We hereafter define $\psi' = -\psi$.
$\rho_Q^{(1)}(t)$ is written as
\beq
 \rho_Q^{(1)}(t) = \frac{\Gamma_0-1}{2\pi\sqrt{\Gamma_0}} I_3(t) ,
\label{eq:rho_Q1-2}
\eeq
where we have defined
\beq
 I_3(t) = \int_0^1  du \frac{1-u^2}{\sqrt{u^2 + \psi'}} \cos (8t u) .
\eeq
Changing the variable by $u' = 1-u$, the integral is arranged into
\barr
 I_3(t) & = & \sqrt{\frac{2}{1+\psi'}}\int_0^1 du'
 \cos \big[ 8t (1-u') \big] \nonumber \\
  && \times \left[ \frac{u'(1 - \frac{u'}{2})}
  {(1-\frac{u'}{1+i\sqrt{\psi'}})(1 -
  \frac{u'}{1-i\sqrt{\psi'}})} \right]^{1/2} \nonumber
\earr
For $0\leq u'\leq 1$, one can expand the term under the square root
in power series of $u'$:
$\left[ \; \cdot \; \right]^{1/2} = \sqrt{u'}(1 + c_1u' + c_2 u'^2 + \cdots)$.
This and the asymptotic expansions of Fresnel's integrals~\cite{abramowitz}
lead to:
\[
 I_3(t) =  - \sqrt{\frac{2}{1+\psi'}} \frac{\sqrt{\pi}}{32\sqrt{2}t^{3/2}}
  \cos \Bigl[8t + \frac{\pi}{4}\Bigr] + \mathcal{O}(t^{-2})
\]
Writing $\psi'$ in terms of $\Gamma_0$, Eq.~\eqref{eq:rho_Q1-2}
is then evaluated as
\beq
 \rho_Q^{(1)}(t) = \frac{1-\Gamma_0}{32\sqrt{\pi}(1+\Gamma_0)\, t^{3/2}}
  \cos \Bigl[8t + \frac{\pi}{4}\Bigr] + \mathcal{O}(t^{-2}) .
\label{eq:rho_Q1-2-final}
\eeq
iii) Case $\Gamma_0 = 1$ and $\Gamma \neq 1$.
Notice that $\psi=u_0^2$. Eq.~\eqref{eq:rho_Q1} is reduced to
\beq
 \rho_Q^{(1)}(t) = \frac{1-\Gamma^2}{4\pi \sqrt{\Gamma}} I_4(t) ,
\label{eq:rho_Q1-3}
\eeq
where we have defined
\beq
 I_4(t) = \int_{u_0}^1 du \frac{\sqrt{1 - u^2}}{u}\cos (t'u) .
\eeq
Changing the variable by $u' = 1-u$, the integral is arranged into
\[
 I_4(t) = \sqrt{2} \int_0^{1-u_0} du' \cos \big[ t'(1-u') \big] 
 \times \Bigg[ \frac{\sqrt{u'(1-\frac{u'}{2})}}{1-u'} \Bigg].
\]
For $0\leq u'\leq 1-u_0$, we perform a power series expansion
in $u'$ as follows:
$\left[ \; \cdot \; \right] = 1 + d_1u' + d_2 u'^2 + \cdots $.
Using this and the asymptotic expansion of Fresnel's integrals~\cite{abramowitz}, 
$I_4(t)$ is evaluated as follows.
\barr
 I_4(t) &=& \sqrt{2} \left\{
  - \frac{\sqrt{1-u_0}}{t} \sin (u_0 t') \right. \nonumber \\
 &&  \times (1+d_1(1-u_0)+d_2(1-u_0)^2+\cdots) \nonumber \\
 && - \left. \frac{\sqrt{\pi}}{2t'^{3/2}}\cos \Bigl[t' +
 \frac{\pi}{4}\Bigr]  +  \mathcal{O}(t'^{-2})
\right\} \nonumber
\earr
The power series of $(1-u_0)$ are summed to yield
\barr
 I_4(t) &=& - \frac{\sqrt{(1-u_0^2)}}{u_0 t'}\sin \big( u_0 t' \big)
  - \sqrt{\frac{\pi}{2}}\frac{1}{t'^{3/2}}
  \cos \Bigl[t' + \frac{\pi}{4}\Bigr]
  \nonumber\\ 
 && + \mathcal{O}(t'^{-2}) . \nonumber
\earr
Writing $u_0$ in terms of $\Gamma$ and substituting $t'=4(1 + \Gamma)t$,
Eq.~\eqref{eq:rho_Q1-3} is evaluated as
\barr
 \rho_Q^{(1)}(t) &=& - \frac{1}{8\pi} \bigg\{
 \frac{1}{t}\sin \big[ 4(1-\Gamma)t \big]
 \label{eq:rho_Q1-3-final} \\
 && +  \frac{\sqrt{\pi}(1-\Gamma)}
 {4\sqrt{2 \Gamma (1+\Gamma)} t^{3/2}} 
 \cos \Bigl[4(1+\Gamma)t + \frac{\pi}{4}\Bigr] \biggr\} \nonumber\\
 && + \mathcal{O}(t^{-2}) . \nonumber
\earr

%++++++++++++++++++++++++++++++++++++++++++++++++++++++++++++++++++++++++++++++++

%++++++++++++++++++++++++++++++++++++++++++++++++++++++++++++++++++++++++++++++++

\end{document}